\documentclass[pdflatex, sn-mathphys-num]{sn-jnl}

\usepackage{graphicx}
\usepackage{multirow}
\usepackage{amsmath,amssymb,amsfonts}
\usepackage{amsthm}
\usepackage{mathrsfs}
\usepackage[title]{appendix}
\usepackage{xcolor}
\usepackage{textcomp}
\usepackage{manyfoot}
\usepackage{listings}
\usepackage{physics}

\title{Impacts of Perfect Fluid Dark Matter on Spacetime Geometry -- the Exponential Metric}
\author*[1]{\fnm{Jan} \sur{Kuncewicz}}\email{kuncewiczjan@gmail.com}
\affil[1]{
    \orgdiv{Institute of Physics},
    \orgname{Maria Curie-Sk\l{}odowska University},
    \orgaddress{\street{pl.~Marii~Curie-Sk\l{}odowskiej~1},
    \city{Lublin}, \postcode{20-031}, \country{Poland}}
}

\abstract{
    Astrophysical observations provide compelling evidence for the existence of
    dark matter, a non-luminous component dominating the universe's mass-energy
    budget. Its gravitational influence is well-established on galactic scales;
    however, dark matter's precise nature and effect on spacetime geometry
    remain open questions. This study investigates modifications to the
    Schwarzschild metric due to the presence of dark matter, modeled as a
    perfect fluid with a specific equation of state. We derive an
    ``exponential'' metric incorporating this dark matter contribution and
    calculate its key characteristics: the event horizon, innermost stable
    circular orbit (ISCO), and photon sphere. Comparing these with Schwarzschild
    predictions reveals distinct deviations dependent on the dark matter
    distribution.  Furthermore, we analyze the orbital velocity profiles derived
    from the exponential metric, demonstrating its potential to explain the
    observed flat rotation curves of galaxies. Our results underscore the
    importance of considering modified metrics in accurately describing
    spacetime near massive objects and provide a theoretical framework for
    further investigations into dark matter's role in galactic dynamics.
}
\keywords{Dark Matter, Black Holes, Modified Gravity, General Relativity, Galaxy Rotation Curves, Exponential Metric}

\begin{document}

\maketitle

\section*{Introduction}

Observations of galaxy rotation curves \cite{rubin1970rotation,
corbelli2000extended}, gravitational lensing \cite{clowe2006direct}, and the
cosmic microwave background radiation \cite{madhavacheril2014current} provide
compelling evidence for the existence of dark matter. However, its fundamental
nature remains unknown. The standard models of galactic dynamics, based on
observed luminous matter and Newtonian gravity, fail to account for the observed
flat rotation curves of galaxies \cite{rubin1970rotation}.  This discrepancy
points to the presence of a non-luminous halo, suggesting that dark matter
contributes significantly to the gravitational potential of galaxies.

Addressing this inconsistency has led to two main avenues of research:  (1)
modifications to General Relativity, exploring alternative theories of gravity
such as MOND (Modified Newtonian Dynamics) \cite{milgrom1983modification,
bekenstein1984does, bekenstein2004relativistic}, or $f(R)$ gravity
\cite{capozziello2012dark,li2012haloes}, and (2) the hypothesis of new, weakly
interacting particles that constitute dark matter \cite{gervais1971field,
wess1974supergauge, peccei1977constraints}.  

These exotic particles are hypothesized by frameworks like supersymmetry
\cite{jungman1996supersymmetric}.  A significant body of work focuses on
analytic halo models, some of which leverage scalar fields. Fay
\cite{fay2004scalar}, for example, employed a Brans-Dicke massless scalar field
in their model, while Matos, Guzmán, and Nuñez \cite{matos2000spherical}
investigated a massless minimally coupled scalar field incorporating a
potential.  In this work, we focus on the second approach and investigate how
the presence of dark matter, modeled as a perfect fluid with a specific equation
of state, affects the spacetime geometry around a central mass within the
framework of General Relativity.  
We find static and spherically symmetric solutions of Einstein equations
that are defined in terms of parameter $\epsilon$ which has been explored
in some works \cite{salgado2003simple,dymnikova2002cosmological,giambo2002anisotropic,kiselev2003quintessence}.

We derive an ``exponential'' metric that modifies the Schwarzschild solution
by incorporating the dark matter contribution. We then calculate the event
horizon, ISCO, and photon sphere for this modified metric.  Comparing these
values to those obtained from the Schwarzschild metric, we identify specific deviations
arising from our dark matter model.
We calculate $r_\epsilon(\epsilon)$ relation given realistic data. From that, we
deduce that those values correspond to small deviations from the Schwarzschild metric.
That result is a good indication of the validity of this model because dark matter contribution
isn't detected on the scale of the Solar system. 

\section*{Derivation of the Exponential Metric}

Our starting point are the Einstein field equations, which relate the curvature
of spacetime to the energy-momentum distribution:

\begin{equation}
G_{\mu\nu} = T_{\mu\nu},
\label{eq:einstein}
\end{equation}
where $G_{\mu\nu}$ is the Einstein tensor, $T_{\mu\nu}$ is the stress-energy
tensor, and the Einstein gravitational constant is equal to 1.

Assuming a spherically symmetric and static spacetime, we use the following metric:

\begin{equation}
\dd s^2 = -e^{\nu(r)} \dd t^2 + e^{\lambda(r)} \dd r^2 + r^2 (\dd\theta^2 + \sin^2\theta \dd\phi^2),
\label{eq:metric}
\end{equation} 
where $\nu(r)$ and $\lambda(r)$ are functions of the radial coordinate, $r$.

We model dark matter as a perfect fluid, meaning the energy-momentum tensor has
a diagonal form of 

\begin{equation}
    T = \text{diag}(T_{tt},\ T_{rr},\ T_{\theta\theta},\ T_{\phi\phi}).
\end{equation}
For this metric, the non-zero components of the Einstein tensor are:
\begin{align}
    G^{t}{}_t &= e^{-\lambda} \qty(\frac{1}{r^2} - \frac{\lambda'}{r}) - \frac{1}{r^2} = T^{t}{}_t,\\
    G^{r}{}_r &= e^{-\lambda} \qty(\frac{1}{r^2} + \frac{\nu'}{r}) - \frac{1}{r^2} = T^{r}{}_r,\\
    G^{\theta}{}_\theta &= \frac{1}{2}e^{-\lambda}\qty(\nu'' + \frac{(\nu')^2}{2} + \frac{\nu'-\lambda'}{r} - \frac{\nu'\lambda'}{2}) = T^{\theta}{}_\theta,\\
    G^{\phi}{}_\phi &=  G^{\theta}{}_\theta = T^{\phi}{}_\phi.
\end{align}
Following \cite{salgado2003simple}, we impose a specific relation between the components:
\begin{equation}
T^{\theta}{}_{\theta} = T^{\phi}{}_{\phi} = T^{t}{}_{t} (1 - \epsilon),
\label{eq:pressure}
\end{equation}
where $\epsilon$ is a constant parameter. 
Our motivation for using this approach is that it is a generalization of power-law-like
solutions in spherically symmetric and static spacetime (examples in Table \ref{tab:ph-met}).
This allows for analytical solutions while introducing non-zero energy-momentum tensor
that can be interpreted as a constant field of dark matter.

Substituting the metric (Eq. \ref{eq:metric}) and the condition on the
energy-momentum tensor (Eq. \ref{eq:pressure}) into the Einstein field equations
(Eq. \ref{eq:einstein}), we obtain a system of differential equations for
$\nu(r)$ and $\lambda(r)$.  Solving these equations, we arrive at two solutions
for $e^{\nu(r)}$, with $\lambda(r) = -\nu(r)$:

\begin{equation}
e^{\nu(r)} = 1 - \frac{r_S}{r} + \frac{r^{2(1-\epsilon)}}{r_\epsilon}
,\ \epsilon\neq\frac32 ;
\label{eq:exp}
\end{equation}
\begin{equation}
e^{\nu(r)} = 1 - \frac{r_S}{r} + \frac{a}{r}\ln\qty(\frac{r}{\abs{a}})
,\ \epsilon=\frac32 ;
\label{eq:log}
\end{equation}
where $r_S = 2M$ is the Schwarzschild radius, and $r_\epsilon$ and $a$ are
integration constants related to the dark matter distribution.  The solution in
Eq. (\ref{eq:log}) has been previously studied \cite{li2012galactic}. Therefore,
we focus on the solution in Eq. (\ref{eq:exp}), which represents the
``exponential'' metric. This metric reduces to the
Schwarzschild solution when $r_\epsilon \to \infty$ and $\epsilon\to 1$.

\begin{table}[ht!]
    \centering
    \begin{tabular*}{\textwidth}{@{\extracolsep{\fill}}c c c c}
        \hline
        Spacetime & Fields & $\epsilon$-value & $r_\epsilon$-value \\ \hline
        Schwarzschild & Vacuum (none) & $\epsilon = 1$ & $ r_\epsilon\to\infty$ \\
        Reissner-N\"ordstrom & Electric field $A_a = -\delta^t_a \frac{Q}{r}$ & $\epsilon = 2 $ &  $r_\epsilon = 1/GQ^2$  \\
        de Sitter/anti-de Sitter & Cosmological constant $\Lambda = \text{const.}$& $\epsilon = 0$ & $r_\epsilon = -3/\Lambda$ \\
        \hline
    \end{tabular*}
    \caption{Example values that generate known metric solutions \cite{salgado2003simple,li2012galactic}}.
    \label{tab:ph-met}
\end{table}

To analyze the motion of test particles, we consider the
geodesic equations. For massive particles, the equations of motion can be
expressed in terms of
\begin{equation}
    \qty(\derivative{r }{\tau})^2 = \mathcal{E}^2 - \qty(\frac{\mathcal{L}^2}{r^2} + 1)g_{tt},
    \label{eq:eom}
\end{equation}
where $\tau$ is the proper time, $\mathcal{E}$ is the conserved energy per unit
mass, and $\mathcal{L}$ is the conserved angular momentum per unit mass. The
effective potential is given by
\begin{equation}
    V_{eff}(r) = 1 - \frac{r_S}{r} + \frac{\mathcal{L }^2}{r^2} - \frac{\mathcal{L}^2 r_S }{r^3} + \frac{r^{2(1-\epsilon)}}{r_\epsilon} + \frac{\mathcal{L }^2r^{2(1-\epsilon)}}{r^2r_\epsilon}.
    \label{eq:veff}
\end{equation}

It is important to note that for the exponential metric, not all values of
$\epsilon$ are physically admissible. The effective potential should satisfy two conditions:
asymptotic flatness $\lim_{r\to\infty} V_{eff}(r) < \infty$.

\section*{Characteristic Values of the Exponential Metric}

We now proceed to calculate the characteristic values for the exponential metric
using techniques based on \cite{shaymatov2021effect}.
Meaning that, we expand our solution for small values around Schwarzschild metric.

\subsection*{Event Horizon}

The event horizon is defined as the radius $r_h$ where $e^{\nu(r_h)} = 0$. Solving this equation yields:

\begin{equation}
r_h = \frac{1}{2}r_{S} + \frac{1}{2}r_{S}\sqrt{1 - \frac{4r_{S}^{2(1-\epsilon)}}{r_\epsilon}}.
\label{eq:horizon}
\end{equation}

\subsection*{Innermost Stable Circular Orbit (ISCO)}

The ISCO corresponds to the minimum of the effective potential for massive test
particles.  It is found by solving $\partial_r V_{eff}(r_{ISCO}) = 0$ and
$\partial^2_r V_{eff}(r_{ISCO}) = 0$ and it is given by

\begin{equation}
\begin{split}
    r_{ISCO} &= 3r_{Sch} + r_{Sch}r^{2(1-\epsilon)}_{Sch}\frac{4\epsilon^2 - 20\epsilon + 15}{r^2_\epsilon} +\\
             &- 4 r_{Sch} r_{Sch}^{2(1-\epsilon)} \frac{(\epsilon - 1)(\epsilon - 2)}{r_\epsilon} +\\
             &+ 4 r_{Sch}^{4(1-\epsilon)}\frac{\epsilon(\epsilon - 1)}{r_\epsilon^2 r_{Sch}},
\end{split}
\end{equation}
expanded up to the order of $\epsilon^2/r_\epsilon^2$ for small values around
the Schwarzschild solutions.

\subsection*{Photon Sphere}

The photon sphere is defined as the radius $r_{ph}$ where photons can follow circular orbits.
Changing Eq. \eqref{eq:eom} to massless particles yields to a new expression for the effective potential.

\begin{equation}
    V_{eff}(r) = \frac{\mathcal{L }^2}{r^2} - \frac{\mathcal{L}^2 r_S }{r^3} + \frac{\mathcal{L }^2r^{2(1-\epsilon)}}{r^2r_\epsilon}.
\end{equation}
Solving $\eval{\dv*{V_{ph}(r)}{r}}_{r_{ph}} = 0$, we find the photon sphere radius
\begin{equation}
r_{ph} = \frac{3}{2}r_{Sch} - r_{Sch}r_{Sch}^{2(1-\epsilon)}\frac{\epsilon}{r_\epsilon}.
\label{eq:photonsphere}
\end{equation}

\section*{Circular Orbits in the Exponential Metric}

To further explore the implications of the exponential metric for galactic dynamics, we analyze the properties of circular orbits.  In the framework of General Relativity, the motion of test particles around a central mass is governed by the geodesic equations. For circular orbits, as observed by a distant observer, the radial velocity must vanish, implying
\begin{equation}
\dv{r}{t} = \dv{r}{\tau}\dv{\tau}{t} = 0 \implies \dv{r}{\tau} = 0,
\label{eq:ass1}
\end{equation}
where $\tau$ is the proper time of the test particle.  Furthermore, stable circular orbits correspond to extrema of the effective potential
\begin{equation}
\dv{V_{eff}(r)}{r} = 0.
\label{eq:ass2}
\end{equation}

By combining Equations (\ref{eq:ass1}) and (\ref{eq:ass2}), and substituting the
explicit form of the effective potential for the exponential metric (Eq.
\ref{eq:veff}), we can derive an expression for the orbital velocity $v(r)$ of a
test particle in a circular orbit
\begin{equation}
    v^2(r) = r_S\qty(\frac{1}{2r} + \frac{r^{2(1-\epsilon)}(1-\epsilon)}{r_\epsilon}).
    \label{eq:vel}
\end{equation}
This expression differs from the standard Schwarzschild and Newtonian result of
$v^2(r) = r_S/2r$ by the presence of an additional term that depends on the dark
matter parameters $\epsilon$ and $r_\epsilon$.  This additional term represents
the contribution of the dark matter halo to the gravitational potential, leading
to a modification of the orbital velocity profile.  The deviation from the
standard $1/r$ dependence becomes more pronounced at larger radii, precisely
where the dark matter halo's influence is expected to be most significant.

\section*{Discussion}

Having established the theoretical framework for our exponential metric, we now
turn to a quantitative analysis of its key characteristics.

\begin{figure}[ht!]
    \centering
    \includegraphics[width=0.6\textwidth]{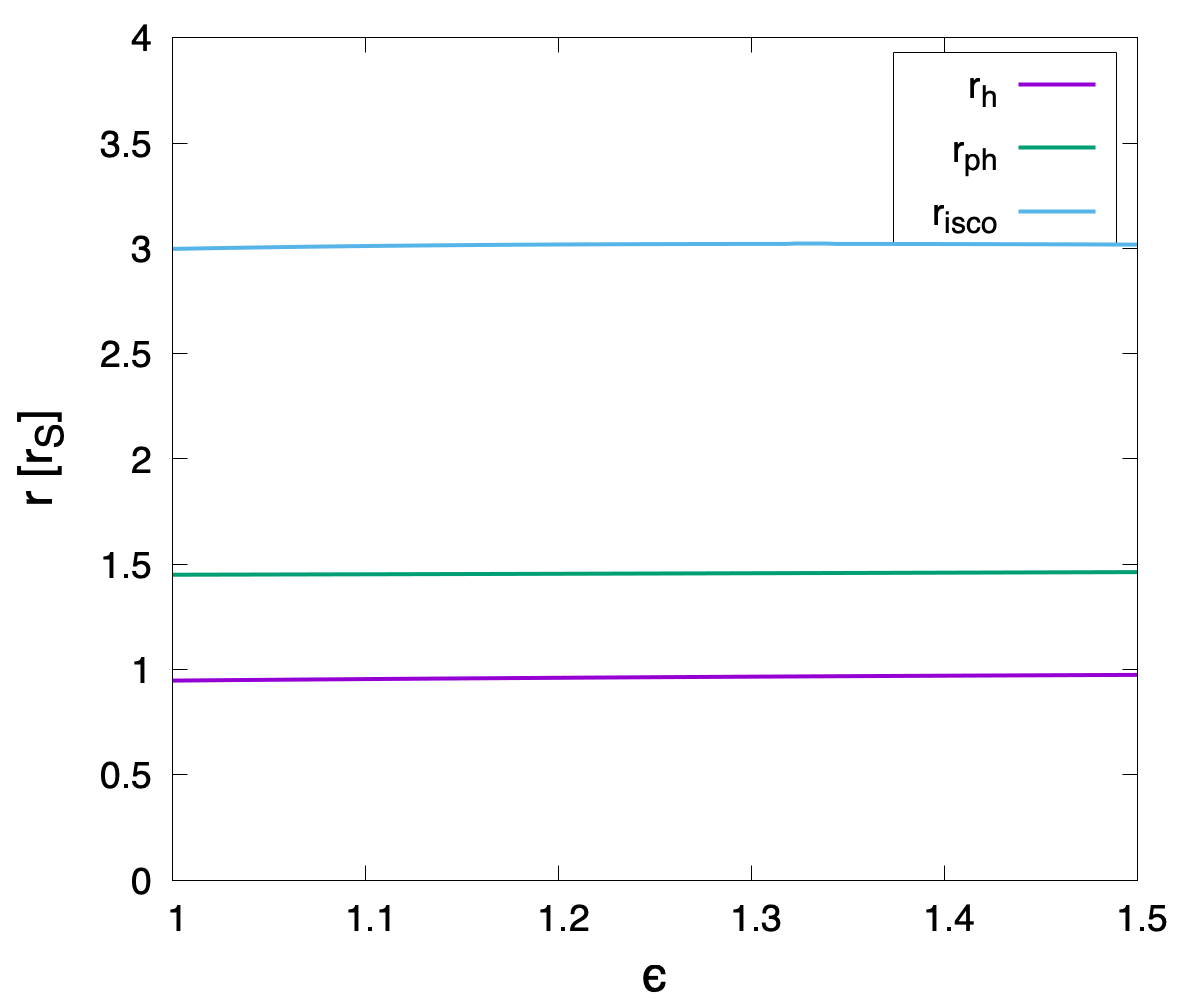}
    \caption{Characteristic radii for the exponential metric as a function of
    the dark matter parameter $\epsilon$ with $r$ in units of $r_S$, with
    $r_\epsilon = 20 r_{Sch}$ fixed: (a) Event horizon radius ($r_h$), (b)
    Innermost stable circular orbit radius ($r_{ISCO}$), and (c) Photon sphere
    radius ($r_{ph}$). }
    \label{fig:r-exp}
\end{figure}

Figure \ref{fig:r-exp} illustrates the variation of $r_h$, along with the ISCO
radius ($r_{ISCO}$) and photon sphere radius ($r_{ph}$), with respect to
$\epsilon$ for a fixed value of $r_\epsilon$.  As evident from the figure,
increasing the dark matter influence, represented by increasing $\epsilon$,
leads to a larger event horizon radius compared to the standard Schwarzschild
value ($r_h = r_{Sch}$ for $r_\epsilon \to\infty$).  This behavior can be attributed to
the additional gravitational pull exerted by the dark matter halo surrounding
the central mass. 

Furthermore, Fig. \ref{fig:r-exp} demonstrates that both the ISCO radius and
photon sphere radius also increase with increasing $\epsilon$.  This expansion
of stable orbital radii is consistent with the presence of a dark matter halo,
which would effectively reduce the gravitational force experienced by test
particles at a given distance from the central mass.

\begin{figure}[ht!]
    \centering
    \includegraphics[width=0.6\textwidth]{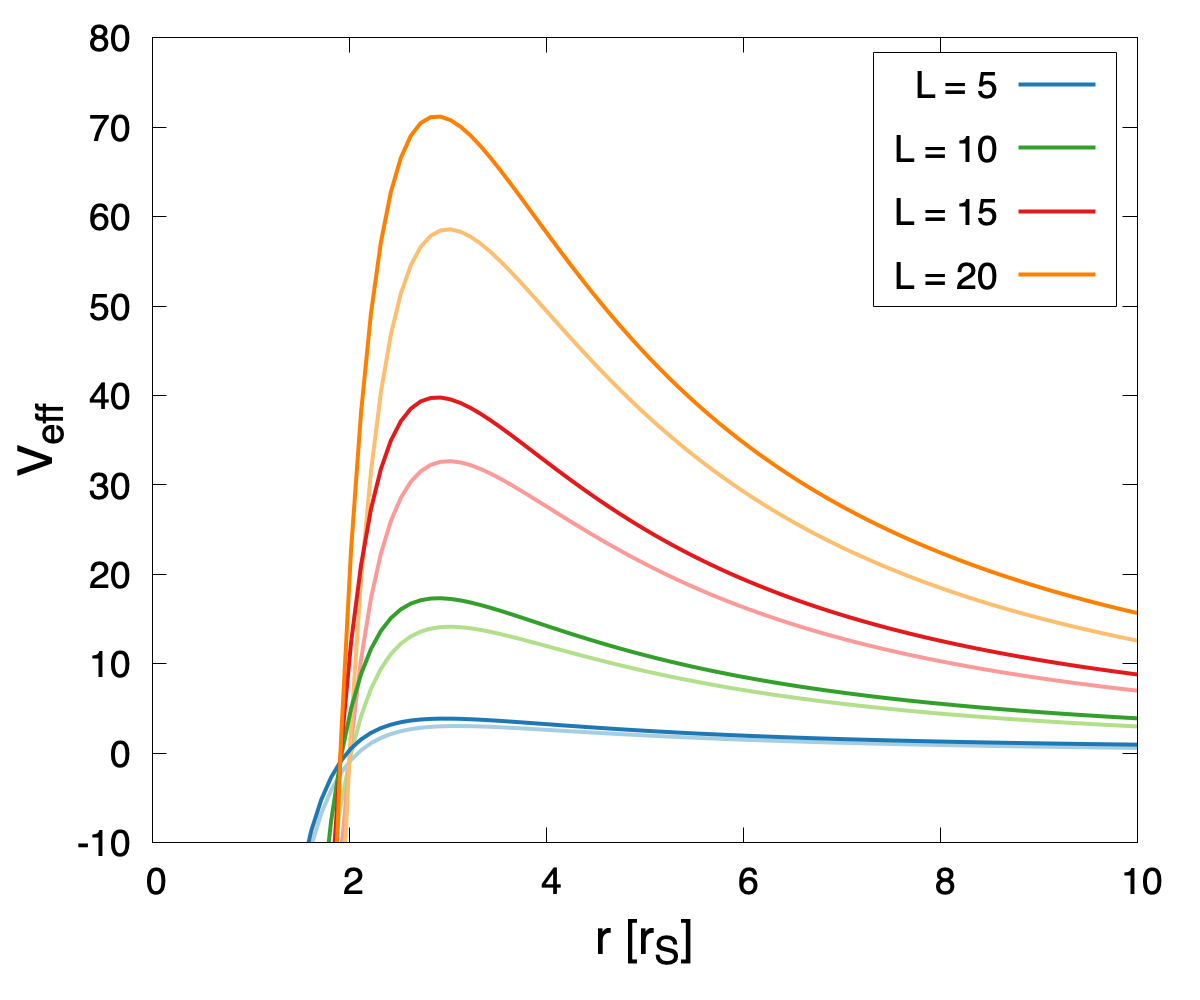}
    \caption{Comparison of the effective potential for the exponential metric
    (darker colors) with the Schwarzschild metric (lighter colors) for different
    values of specific angular momentum ($\mathcal{L}$) with $\epsilon = 1.2$,
    $r_\epsilon = 20r_{Sch}$ and $r$ in units of $r_S$.}
    \label{fig:exp-comp}
\end{figure}

\begin{figure}[ht!]
    \centering
    \includegraphics[width=0.6\textwidth]{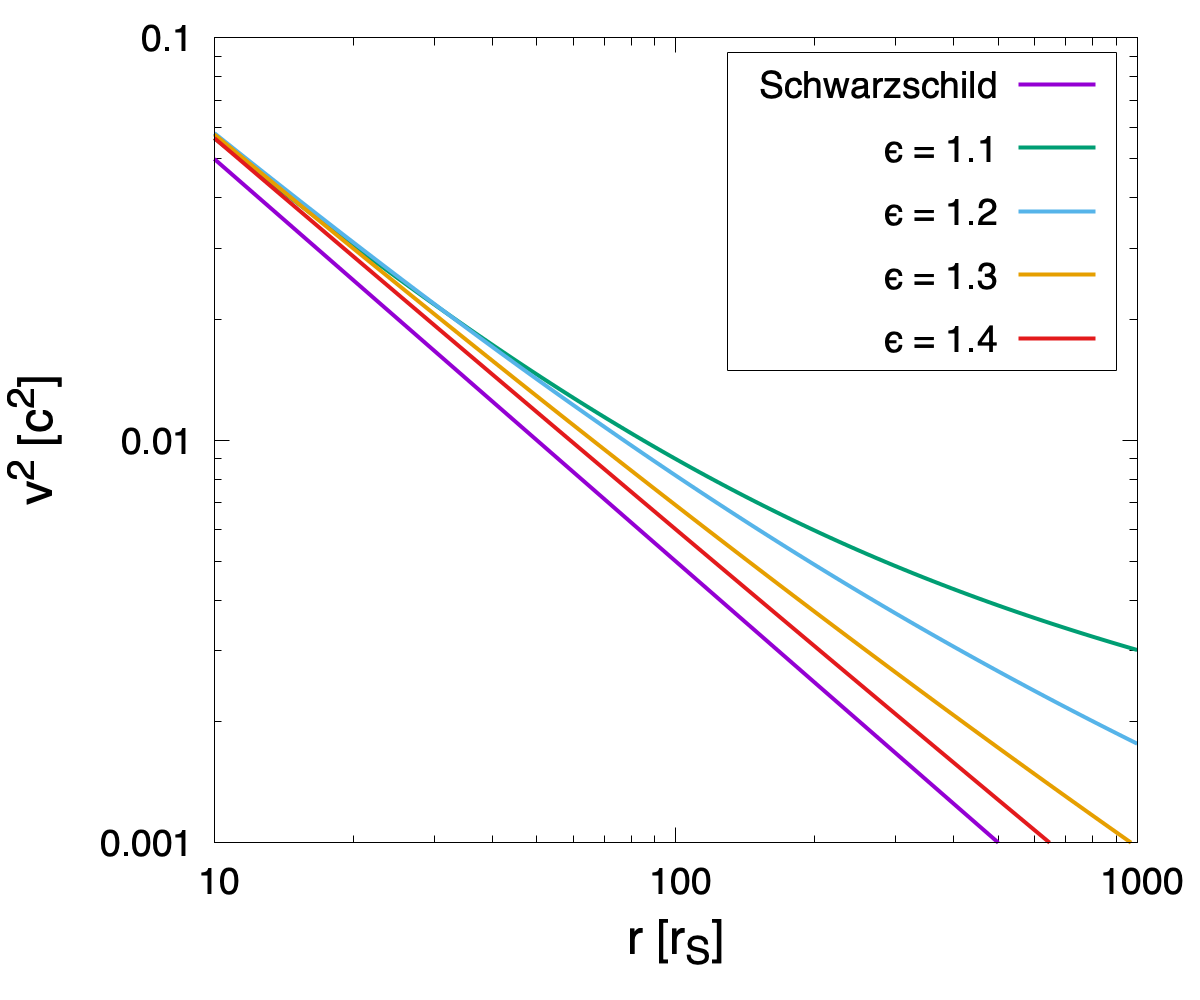}
    \caption{Logarithmic plot of $v^2(r)$ for the Schwarzschild metric and the
    exponential metric (Eq. \eqref{eq:vel}) with $r_\epsilon = -10r_S$ and $v^2$ in units of $c^2$.}
    \label{fig:vel-exp}
\end{figure} 

To further highlight the impact of dark matter on the spacetime geometry, we
compare the effective potential for the exponential metric to that of the
Schwarzschild metric in Fig. \ref{fig:exp-comp}.  The presence of dark matter,
as incorporated in the exponential metric, leads to a shallower potential well
compared to the Schwarzschild case. This difference in potential is particularly
pronounced at larger radii, indicating that the gravitational influence of dark
matter extends beyond the central mass distribution.

The shallower potential well in the exponential metric has significant
implications for the dynamics of test particles. For a given value of specific
angular momentum, test particles can exist in stable orbits at larger radii
compared to the Schwarzschild case. This result provides a potential explanation
for the observed flat rotation curves of galaxies.  If dark matter is indeed
distributed in a halo surrounding a galaxy, its gravitational influence, as
captured by the exponential metric, could account for the higher-than-expected
orbital velocities of stars and gas in the outer regions of galaxies. 

Figure \ref{fig:vel-exp} depicts the orbital velocity profiles.  The addition of
dark matter in the exponential metric leads to higher circular velocities at
larger radii, with a slower decay compared to the Schwarzschild case.  This
result further supports the potential of the exponential metric to address the
galaxy rotation curve problem.  For instance, the $\epsilon=1.1$ case exhibits a
slower velocity decline, reflecting the non-constant contribution of the dark
matter term in Eq. (\ref{eq:vel}). This behavior warrants further investigation,
such as hydrodynamical simulations incorporating the modified energy-momentum
tensor, to confirm its role in galactic dynamics.

\section*{Estimation of parameters}
The theoretical framework developed for the exponential metric necessitates a
robust methodology for estimating its free parameters, $\epsilon$ and $r_\epsilon$.
These parameters encode the influence of dark matter on the spacetime geometry
and, consequently, on the dynamics of test particles.  Therefore, a precise
determination of these parameters is crucial for validating the model against
observational data.

Our rotation curves (Figure \ref{fig:vel-exp}) take into account only central
supermassive black hole without surrounding galactic matter, which means that we
cannot calculate precise velocities from Equation \eqref{eq:vel}.  We should be
able to estimate the ``excess'' velocity, meaning that which cannot be explained
from classical estimates without dark matter.  Our assumption agrees with the
fact that estimated dark matter halos are connected with supermassive black
holes in the center of galaxies \cite{baes2003observational}.

To illustrate the parameter estimation procedure, we consider a representative
scenario where the central mass is of the order $M \sim 10^7 M_\odot$. We
analyze the orbital velocity at a radial distance of $r = 50$ kpc, where the
velocity is expected to be in the order of $v \sim 200$~km~s$^{-1}$, consistent with
observations of many well-measured galactic rotation curves
\cite{sanders1996published}.

\begin{figure}[ht!]
    \centering
    \includegraphics[width=0.6\textwidth]{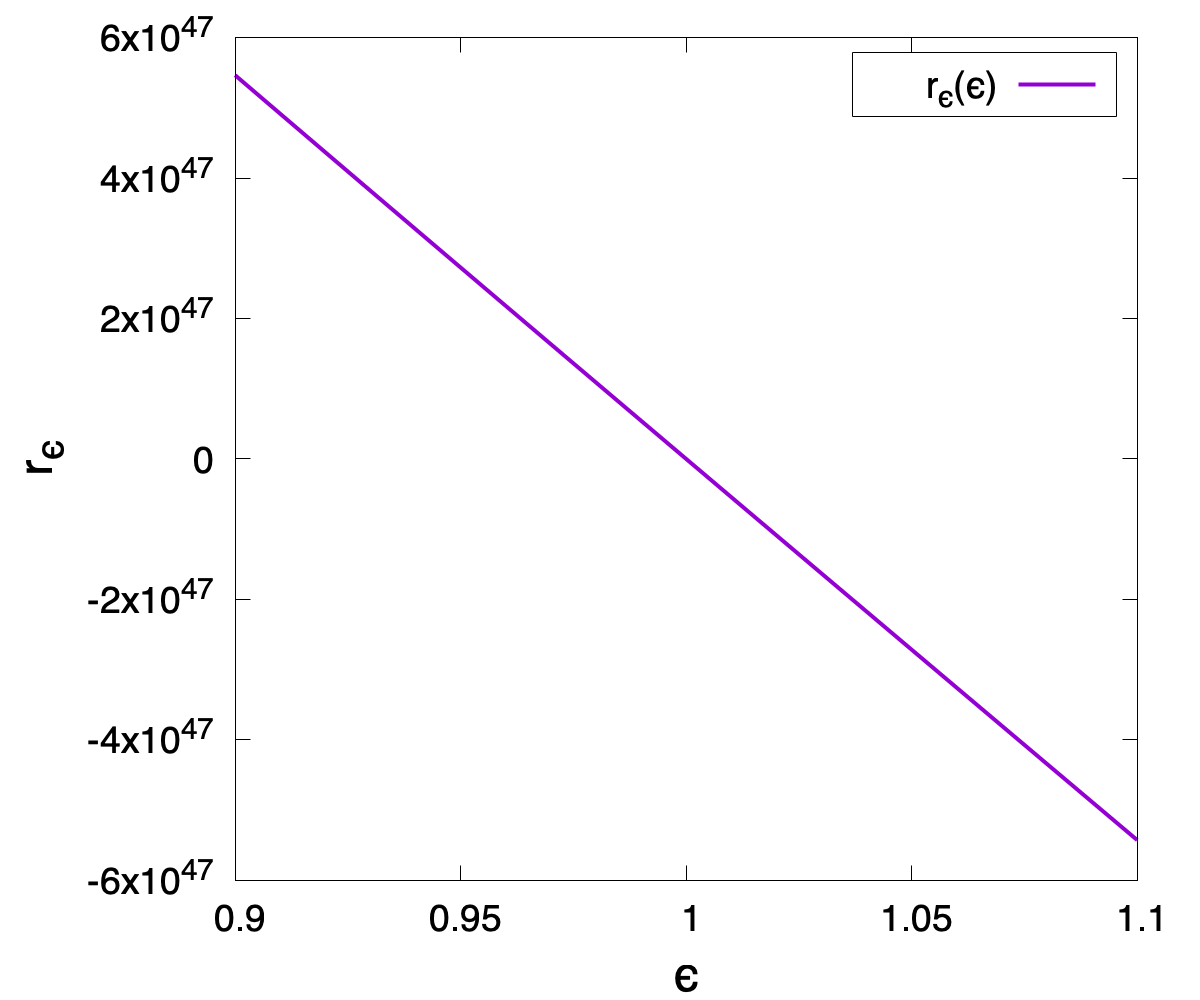}
    \caption{Relation $r_\epsilon(\epsilon)$ for $M = 10^7$~M$_\odot$, $r = 50$~kpc and $v=200$ km s$^{-1}$.} 
    \label{fig:vel-real}
\end{figure}

Solving for $r_\epsilon$ as a function of $\epsilon$, denoted as $r_\epsilon(\epsilon)$,
allows us to explore the parameter space and identify viable combinations of
$\epsilon$ and $r_e$ that reproduce the observed rotation curves. Figure \ref{fig:vel-real}
presents an example of such a relationship, calculated for $M=10^7
M_\odot$, $r = 50$ kpc, and $v = 200$ km s$^{-1}$ within the range $\epsilon \in
[0.9, 1.1]$, excluding $\epsilon = 1$ (as this corresponds to the Schwarzschild
solution).

As illustrated in Figure \ref{fig:vel-real}, values of $\epsilon < 1$ lead to
large $r_e$ values.  This finding is consistent with the expectation that small
perturbations to the Schwarzschild metric, represented by small deviations of
$\epsilon$ from unity, require a larger dark matter contribution, characterized
by a larger $r_e$. This analysis provides a framework for constraining the
parameters of the exponential metric using observational data.  Future work will
focus on extending this analysis to a broader range of galactic rotation curves
and exploring the implications for dark matter halo profiles.

\section*{Conclusion}

We have investigated the modifications to the spacetime geometry
around a central mass due to the presence of dark matter. Then we derived an
exponential metric by incorporating the contribution of dark matter, modeled
as a perfect fluid with a specific condition on energy-momentum tensor, into the Einstein field
equations.

Our analysis of the exponential metric revealed several key findings:
\begin{enumerate}
    \item Larger event horizon, ISCO, and photon sphere radii compared to the Schwarzschild metric, indicating dark matter's extended gravitational influence.
    \item A shallower effective potential, particularly at larger radii, suggesting a dark matter halo around the central mass.
    \item Circular orbital velocities that decay slower at larger radii than predicted by the Schwarzschild metric. This slower decay offers a potential explanation for the observed flat rotation curves of galaxies. 
    \item In our model $r_\epsilon$ might be understood as a characteristic length of dark matter in the galaxy and $\epsilon$ as a EOS proportionality constant as it appears in the Eq. \eqref{eq:pressure}.
\end{enumerate}

These findings underscore the importance of considering modified metrics that incorporate dark matter when studying the gravitational fields of massive objects. While our study has focused on a specific dark matter model and a static, spherically symmetric spacetime, future research could explore more complex scenarios, including:
\begin{itemize}
    \item Investigating the exponential metric in the context of rotating black holes (Kerr metric).
    \item Exploring different dark matter equations of state and their impact on the spacetime geometry.
    \item Conducting numerical simulations to study the dynamics of matter in the presence of the exponential metric.
    \item Studying a wide range of data to uncover detailed ranges of $\epsilon$ and $r_\epsilon$.
\end{itemize}
It is worth noting that some work has been done on Kerr \cite{hou2018rotating}
and Bardeen \cite{narzilloev2020dynamics,zhang2021bardeen} black holes, although only
in terms of logarithmic metric. Future work will be done to calculate parameters in
the exponential metric regime in Kerr and Reissner-N\"orsdrom black holes. 

\section*{Acknowledgements}
I would like to thank M. Rogatko and P. Verma for helpful comments on manuscript and many valuable discussions.

\bibliography{refs}
\end{document}